\documentclass[RNAAS]{aastex631}
\usepackage{amsmath}
\usepackage{bm}
\usepackage{physics}

\shorttitle{State Space GPs}
\shortauthors{Jord\'an et al.}

\graphicspath{{./}{figures/}}

\begin{document}

\title{State-space representation of Mat\'ern and Damped Simple Harmonic Oscillator Gaussian processes}

\author{Andr\'es Jord\'an}
\author{Susana Eyheramendy}
\affiliation{Facultad de Ingenier\'ia y Ciencias, Universidad Adolfo Ib\'a\~nez, Av.\ Diagonal las Torres 2640, Pe\~{n}alol\'{e}n, Chile}
\affiliation{Millennium Institute for Astrophysics, Chile}

\author{Johannes Buchner}
\affiliation{Max-Planck-Institut f\"ur Extraterrestrische Physik,
Giessenbachstrasse, 85748 Garching, Germany}

\begin{abstract}
Gaussian processes (GPs) are used widely in the analysis of astronomical time series. GPs with rational spectral densities have state-space representations which allow $\mathcal{O}(n)$ evaluation of the likelihood. We calculate analytic state space representations for the damped simple harmonic oscillator and the Mat\'ern 1/2, 3/2 and 5/2 processes.
\end{abstract}


\section{Introduction}
\label{sec:intro}

Gaussian processes are now widely used in the analysis of astronomical time series. Efficient algorithms have been devised that allow likelihood evaluation in $\mathcal{O}(n)$ operations, where $n$ is the number of observations \citep[see][and references therein]{dfm:2017:celerite}. Linear Gaussian state space models allow likelihood evaluation in $\mathcal{O}(n)$ operations via the Kalman filter \citep{kalman:1960}. In this note we present analytic linear Gaussian state space model representations for the popular Mat\'ern and Damped Simple Harmonic Oscillator (DSHO) Gaussian processes, offering an alternative route to efficient likelihood evaluation for these processes. This representation can offer advantages in terms of numerical stability for the Matérn case (D. Foreman-Mackey 2021, personal communication).

There is a correspondence between a stationary Gaussian Process (GP) over $\mathbb{R}$ with  kernel $k(\cdot)$ and power spectral density $S(\omega)$ that is a rational function of $\omega$, i.e. $S(\omega) \propto p(\omega)/q(\omega)$ with $p$ and $q$ polynomials, and  the prior implied by an order $p$ linear stochastic differential equation given by

\begin{align}
\dv[p]{f}{t} + & a_{1}\dv[p-1]{f}{t}+\cdots+a_p\dv{f}{t} \nonumber\\
& = b_0\dv{W}{t}+b_1\dv[2]{W}{t}+\cdots+b_q\dv[q+1]{W}{t},
\label{eq:sde}
 \end{align}
 where $W$ is a Wiener process (Brownian motion) with variance  $\sigma^2_w$ \citep{brockwell_davis:2013, hartikainen:2010:kalman, saatcci:2012}. Equation~\ref{eq:sde} is equivalent to a state space model with the following {\em state} equation

 \begin{equation}
 \dv{\bm{x}}{t} = {\bf A} {\bm x}(t) + {\bm L}dW(t)
 \label{eq:vec_markov}
 \end{equation}

\noindent where ${\bm x} = (f(t),\dv{f}{t},\ldots,\dv[p-1]{f}{t})$,
${\bm L} = (0,0,\ldots,1)^\intercal$ and

\[
{\mathbf A} = \left(
\begin{array}{ccccc}
0 & 1 & \cdots & 0 & 0\\
\vdots & \ddots & \cdots & \ddots & \vdots\\
0 & 0 & \cdots & 1 & 0\\
-a_p & -a_{p-1} & \cdots & -a_{2} & -a_{1}\\
\end{array}
\right),
\]

\noindent and the following {\em measurement} equation

\[
y = \bm{b}^{\intercal}\bm{x}
\]

\noindent where $\bm{b}=(b_0, b_1,\ldots,b_{p-1})^\intercal$ and $b_j=0$ for $j>q$.  The observed process $y$ would in the astronomical context used to model a time series $y(t)$ observed at times  $\{t_i\}_{i=1}^n$ with associated measurement errors $\{\sigma_i\}_{i=1}^n$.

 If $\bm{x}(t)$ is stationary the process $y$ defined by the above state space model can be shown \citep[e.g.][]{brockwell:2009} to be a zero-mean Gaussian CARMA($p,q$) model with  power spectral density

 \begin{equation}
 S(\omega) = \frac{\sigma_w^2}{2\pi}\left|\frac{b(i\omega)}{a(i \omega)}\right|^2
 \label{eq:carma_psd}
 \end{equation}
and variance $\bm{\Sigma}$ that satisfies

\begin{equation}
\dv{ \bm{\Sigma}(t) }{t} = \bm{A \Sigma}(t) + \bm{\Sigma A^{\intercal}} + \sigma_w^2 \bm{L L^{\intercal}}.
\label{eq:ode_sigma}
\end{equation}

 We can express the process given by equation~\ref{eq:vec_markov} as a linear Gaussian state space model observed at a set of input points $\{\bm{x}(t_i)\}$

\begin{align}
p(\bm{x}(t_i) | \bm{x}(t_{i-1})) &=  N(\bm{\Phi}_{i-1}\bm{x}(t_{i-1}),\bm{Q}_{i-1}) \nonumber\\
p(y(t_i) | \bm{x}(t_i)) &=  N( \bm{b}^\intercal \bm{x}(t_i), \sigma^2_n)
\label{eq:sm_dynamics}
\end{align}

\noindent where $\delta_i \equiv t_i-t_{i-1}$, $\bm{\Phi}_{i-1} \equiv \exp(\bm{A}\delta_i)$ and

\begin{equation}
\bm{Q}_{i-1} = \sigma^2_w \int_{0}^{\delta_i} (\delta_i - h)^2 \bm{c}_{i-1}  \bm{c}_{i-1}^{\intercal} \,dh,
\label{eq:Q}
\end{equation}

\noindent where $\sigma^2_n$ is the measurement error, $\bm{c}_{i-1}$ is the last column of the matrix $\exp(\bm{A}(\delta_i-h))$
and $\bm{H}$ is a projection matrix that selects the first component of the vector, which is the process in our state space model. With all the conditional probabilities of our process determined, we can perform inference using a filtering algorithm.

A {\em filtering algorithm} computes the conditional probabilities of the states $\bm{x}_k$ given all previous measurements $\{\bm{y}_1, \bm{y}_2, \ldots, \bm{y}_k\} \equiv \bm{y}_{1:k}$ and a set of parameters $\bm{\theta}$, i.e. $p(\bm{x}_k| \bm{y}_{1:k}, \bm{\theta})$. For the linear Gaussian case
this can done in $\mathcal{O}(n)$ operations with a set of recursive equations known as the Kalman filter, which are given by\footnote{An excellent introduction to Bayesian filtering is given by \citet{sarkka:2013}, from which we take our notation.}

\begin{align}
p(\bm{x}_k | \bm{y}_{1:k-1}, \bm{\theta}) &=  N(\bm{\mu}_k^-, \bm{P}_k^-) \nonumber\\
p(\bm{x}_k | \bm{y}_{1:k}, \bm{\theta}) &=  N(\bm{\mu}_k, \bm{P}_k) \nonumber\\
p(\bm{y}_k | \bm{y}_{1:k-1}, \bm{\theta}) &=  N(\bm{H}_k\bm{\mu}_k^-,\bm{S}_k).
\label{eq:kalman}
\end{align}

\noindent The mean and covariance matrices of the distributions above are obtained with the Kalman filter prediction and update steps (note that the matrices can depend on $\bm{\theta}$ although we have not written explicitly that dependence). The prediction step is given by

\begin{align}
\bm{\mu}_k^- &=  \bm{\Phi}_{k-1}\bm{\mu}_{k-1} \nonumber\\
\bm{P}_k^- &=  \bm{\Phi}_{k-1}\bm{P}_{k-1}\bm{\Phi}_{k-1}^{\intercal} + \bm{Q}_{k-1},
\label{eq:kalman_update}
\end{align}

\noindent and the update step by

\begin{align}
\bm{v}_k &=  \bm{y}_k - \bm{H}_k\bm{\mu}_k^-\nonumber\\
\bm{S}_k &=  \bm{H}_k\bm{P}_k^-\bm{H}_k^{\intercal} + \bm{R}_k \nonumber\\
\bm{K}_k &=  \bm{P}_k^-\bm{H}_k^{\intercal}\bm{S}_k^{-1}\nonumber\\
\bm{\mu}_k &=  \bm{\mu}_k^- + \bm{K}_k \bm{v_k} \nonumber\\
\bm{P}_k &=  \bm{P}_k^- - \bm{K}_k\bm{S}_k\bm{K}_k^{\intercal}.
\label{eq:kalman_predict}
\end{align}

The recursion is initialized with a prior mean $\bm{\mu}_0$ and variance $\bm{P}_0$.  To learn the parameters of a GP that has a corresponding state space representation we are interested in the marginal likelihood which can be computed recursively as

\begin{equation}
p(\bm{y}_{1:n} | \bm{\theta}) = \prod_{k=1}^n p(\bm{y}_k | \bm{y}_{1:k-1}, \bm{\theta}),
\label{eq:pred_error}
\end{equation}

\noindent where we define $p(\bm{y}_1 | \bm{y}_{0:1}) \equiv p(\bm{y}_1)$. The terms in the product are part of the output of the Kalman filter (see equation~\ref{eq:kalman}). Thus, the marginal likelihood can be calculated in $\mathcal{O}(n)$ operations. Armed with the likelihood, we can sample the posterior using a variety of techniques. We note that the use of the Kalman filter to evaluate the likelihood of a CARMA($p,q$) model in the astronomical context has been presented before in \citet{kelly:2014}. In their work the likelihood for a CARMA($p,q$) process is evaluated numerically using the algorithm of \citet{jones:1990} which relies on the diagonalization of the $\bm{A}$ matrix  using the roots of the $a(z)$ polynomial. In this note we provide analytic expressions for the $\bm{\Phi}$ and $\bm{Q}$ matrices for some GP kernels of wide astronomical application.

\section{State Space Models Corresponding to GP Kernels of Interest}

\subsection{The Mat\'ern family}
\label{ssec:matern}

One of the most widely used kernels is the Mat\'ern family, which are indexed by $\nu$ and have an associated spectral density for a one-dimensional input given by \citep{rasmussen:2006}

\begin{equation}
S(\omega) = \sigma^2_f\frac{2\pi^{1/2}\Gamma(\nu+1/2)\lambda^{2\nu}}{\Gamma(\nu)(\lambda^2 + \omega^2)^{\nu+1/2}}
\label{eq:matern_psd}
\end{equation}

\noindent and is thus of the form of equation~(\ref{eq:carma_psd}) for $\nu=n+1/2$, $n\in \mathbb{N}$. More precisely, we have $a(z)=(\lambda+z)^{n+1}$ and $b(z)=1$.
As an example, for the popular $\nu=3/2$ kernel the corresponding stochastic differential equation is

\begin{equation}
\dv{\mathbf{x}(t)}{t} = \left(
\begin{array}{cc}
0 & 1\\
-\lambda^2 & -2\lambda\\
\end{array}
\right)\mathbf{x}(t) + \left(
\begin{array}{c}
0 \\
1\\
\end{array}
\right)dW(t),
\end{equation}

\noindent where $\bm{x(t)} \equiv (x(t),\dv{x(t)}{t})$ and the variance of $dW$ is given in terms of $\sigma^2_f$ and $\lambda$ by

\[
\sigma_w^2 = \sigma_f^2\frac{4\pi^{3/2}\Gamma(2)\lambda^{3}}{\Gamma(3/2)}
\]

\noindent in order that the power spectra density of the process corresponds to that of Equation~\ref{eq:matern_psd} for $\nu=3/2$.

For the Mat\'ern family with $\nu=n+1/2$ the matrices $\bm{\Phi}_{i-1}$ and $\bm{Q}_{i-1}$ defined in equation~(\ref{eq:Q}) can be computed analytically using the Laplace transform and a symbolic mathematics package. We provide the expressions of these matrices for a set of values of $\nu$ below.

\subsection{Stochastically driven Damped Simple Harmonic Oscillator}
\label{ssec:dsho}

\citet{dfm:2017:celerite} proposed the stochastic process given by the following stochastic differential equation (SDE) as a physically motivated process to describe stellar photometric variations

\begin{equation}
\dv[2]{f}{t} + \frac{\omega_0}{Q}\dv{f}{t} + \omega_0^2f(t) = dW(t).
\label{eq:DSHO}
\end{equation}

The corresponding power spectral density is given by

\begin{equation}
S(\omega) = \frac{\sigma^2_w}{(\omega^2-\omega_0^2)^2 + \omega^2\omega_0^2/Q^2}
\label{eq:DSHO_psd}
\end{equation}

\noindent where $\sigma^2_w$ is the spectral density of the white noise process $dW$.
In terms of the power at $\omega_0$ we have that $\sigma^2_w=S(\omega_0)\omega_0^4/Q^2$, and in terms of the total power $S_{\rm tot}$ we have\footnote{This follows from $\int^\infty_{-\infty}\,S(\omega)\,d\omega = \frac{\sigma_w^2 Q}{2\omega_0^3}$ which is obtained using  $\int^\infty_{-\infty}\frac{1}{(a^2-z^2)^2 + b^2z^2}dz = \frac{\pi}{a^2b}.$} that $\sigma^2_w=2S_{\rm tot}\omega_0^3/Q$.

From Equation~\ref{eq:DSHO} we see that the DSHO is equivalent to a CARMA(2,0) process with $a(z)=w_0^2 + (w_0/Q)z + z^2$ and $b(z)=1$.
The corresponding vectorial Markovian SDE is

\begin{equation}
\dv{\mathbf{x}(t)}{t} = \left(
\begin{array}{cc}
0 & 1 \\
-\omega_0^2 & -\omega_0/Q \\
\end{array}
\right)\mathbf{x}(t) + \left(
\begin{array}{c}
0 \\
1\\
\end{array}
\right)dW(t).
\end{equation}

\noindent where $\mathbf{x(t)} \equiv (x(t),\dv{x(t)}{t})$. Just as the case of the Mat\'ern family, the matrices $\bm{\Phi}_{i-1}$ and $\bm{Q}_{i-1}$ defined in equation~(\ref{eq:Q}) can be computed analytically. Note that when $Q=1/2$ this process reduces to a Mat\'ern 3/2 process with $\lambda=\omega_0$.

\section{Calculation of Matrices needed for Likelihood Calculation}

We follow the procedure described in \citet{saatcci:2012} to calculate analytic expressions for the matrices needed for the likelihood calculation for the Mat\'ern and DSHO processes. The matrices $\bm{\Phi}$ are calculated using the  identity

\begin{equation}
\exp(\bm{A}t) = \mathcal{L}^{-1}\{(s\bm{I} - \bm{A})^{-1}\},
\end{equation}

\noindent where $\mathcal{L}^{-1}$ denotes the inverse Laplace transform and $\bm{I}$ is the identity matrix. In the case of the Mat\'ern family of processes, the $\bm{Q}$ matrices can be computed analytically via the integral given in Equation~\ref{eq:Q}. In the case of the DSHO, we use the matrix fraction decomposition. If we define matrices $\bm{C}$ and $\bm{D}$ such that $\bm{\Sigma}=\bm{C}\bm{D}^{-1}$, then equation~\ref{eq:ode_sigma} is satisfied when $\bm{C}$ and $\bm{D}$ take the form

\begin{equation}
\left(
\begin{array}{c}
\bm{C}(t)\\
\bm{D}(t) \\
\end{array}
\right) = \exp\left\{\left(
\begin{array}{cc}
\bm{A} & \bm{L L^{\intercal}}\\
0 & -\bm{A}^{\intercal} \\
\end{array}
\right)t\right\}
\left(
\begin{array}{c}
\bm{C}(0)\\
\bm{D}(0) \\
\end{array}\right),
\label{eq:CD}
\end{equation}

\noindent and $\bm{\Sigma}(t) = \bm{C}(t)\bm{D}(t)^{-1}$. To calculate the $\bm{Q}$ matrices for the DSHO we use equation~\ref{eq:CD} with $\bm{C}(0)=0$ and $\bm{D}(0)=\bm{I}$, calculate the matrix exponential using the inverse Laplace transform, and then set $\bm{Q}_{i-1}=\bm{\Sigma}(\delta_i)$. Note that  we set $\sigma^2_w=1$ to reduce the notational clutter, $\bm{Q}$ matrices presented here should be multiplied by the actual value of this variance when using them. We also denote $\omega_0$ for the damped simple harmonic oscillator simply by $\omega$.

\subsection{Mat\'ern $\nu=1/2$ (Exponential or Ohrstein-Uhlenbeck process)}

\begin{equation}
\bm{\Phi}_{i-1} = \exp(-\lambda\delta_i)
\end{equation}

\begin{equation}
\bm{Q}_{i-1} = -\frac{\exp(- 2\delta_i\lambda) - 1}{2\lambda}
\end{equation}

\subsection{Mat\'ern $\nu=3/2$}

\begin{equation}
\bm{\Phi}_{i-1} = \exp(-\lambda \delta_i)
\left(
\begin{array}{cc}
\lambda\delta_i + 1 & \delta_i \\
-{\lambda}^2 \delta_i   & (1 - \lambda\delta_i)
\end{array}
\right)
\end{equation}

\begin{equation}
\bm{Q}_{i-1} = \frac{1}{4\lambda^2}
\left(
\begin{array}{cc} \lambda^{-1}[1 - \exp(-2\delta\lambda)(2\delta^2 \lambda^2 + 2\delta\lambda + 1)] & 2\delta^2\lambda^2\exp(-2\delta\lambda) \\
2\delta^2\lambda^2\exp(-2\delta\lambda) & \lambda[1 - \exp(-2\delta\lambda)(2\delta^2\lambda^2 - 2\delta\lambda + 1)]
\end{array}
\right)
\end{equation}

\subsection{Mat\'ern $\nu=5/2$}

\begin{equation}
\bm{\Phi}_{i-1} = \exp(-\lambda \delta_i)
\left(
\begin{array}{ccc} \frac{\left(\lambda^2\delta^2 + 2 \lambda\delta + 2\right)}{2} & \delta \left(\lambda\delta + 1\right) & \frac{\delta^2}{2}\\
 -\frac{{\lambda}^3 \delta^2}{2} & \left( -\lambda^2\delta^2 + \lambda\delta + 1\right) & -\frac{\delta\left(\lambda \delta - 2\right)}{2}\\
  \frac{{\lambda}^3\delta\left(\lambda\delta - 2\right)}{2} & \lambda^2\delta\left(\lambda\delta - 3\right) & \frac{\left(\lambda^2\delta^2 - 4\lambda\delta + 2\right)}{2}
\end{array}
\right)
\end{equation}

\begin{align}
\bm{Q}_{i-1} = & \left(
\begin{array}{ccc}
\frac{3}{16\, {\lambda}^5}  & 0  & \frac{1}{16\, {\lambda}^3}\\
 0 & \frac{1}{16\, {\lambda}^3}  & 0  \\
 \frac{1}{16\, {\lambda}^3} & 0  & \frac{3}{16\, \lambda}
\end{array}
\right) + \exp(-\lambda \delta_i)\times
\\
&  \left(
\begin{array}{ccc}
- \frac{\left(2\, {\delta}^4\, {\lambda}^4 + 4\, {\delta}^3\, {\lambda}^3 + 6\, {\delta}^2\, {\lambda}^2 + 6\, \delta\, \lambda + 3\right)}{16\, {\lambda}^5} & \frac{{\delta}^4}{8} & \frac{\left(2\, \delta\, \lambda  + 2\, {\delta}^2\, {\lambda}^2 + 4\, {\delta}^3\, {\lambda}^3 - 2\, {\delta}^4\, {\lambda}^4 + 1\right)}{16\, \lambda^3}\\
 \frac{{\delta}^4}{8} &  - \frac{\left(2\, {\delta}^4\, {\lambda}^4 - 4\, {\delta}^3\, {\lambda}^3 + 2\, {\delta}^2\, {\lambda}^2 + 2\, \delta\, \lambda + 1\right)}{16\, {\lambda}^3} & \frac{{\delta}^2{\left(\delta\, \lambda - 2\right)}^2}{8}\\
  \frac{\left(2\, \delta\, \lambda + 2\, {\delta}^2\, {\lambda}^2 + 4\, {\delta}^3\, {\lambda}^3 - 2\, {\delta}^4\, {\lambda}^4 + 1\right)}{16\, {\lambda}^3} & \frac{{\delta}^2{\left(\delta\, \lambda - 2\right)}^2}{8} & - \frac{\left(2\, {\delta}^4\, {\lambda}^4 - 12\, {\delta}^3\, {\lambda}^3 + 22\, {\delta}^2\, {\lambda}^2 - 10\, \delta\, \lambda + 3\right)}{16\, \lambda}
\end{array}
\right)
\end{align}

\subsection{DSHO with parameters $\omega$ and $Q$. Case $Q\geq1/2$.}

Define $\beta=\sqrt{Q^2-1/4}$.

\begin{equation}
\bm{\Phi}_{i-1} = \exp(-\omega \delta_i/2Q)
\left(
\begin{array}{cc}
 \cos(\omega\beta\delta_i/Q) + \sin(\omega\beta\delta_i/Q)/2\beta &
 Q\sin(\omega\beta\delta_i/Q)/\omega\beta\\
  -Q\omega\sin(\omega\delta_i\beta/Q)/\beta &
 \cos(\omega\delta_i\beta/Q) - \sin(\omega\delta_i\beta/Q)/2\beta \\
\end{array}
\right)
\end{equation}

\begin{align}
\bm{Q}_{i-1} =&
Q\exp(-\omega\delta_i/Q)\times\\
& \left(
\begin{array}{cc}
 \frac{(\cos(2\omega\delta_i\beta/Q)-1) - 2\beta \sin(2\omega\delta_i\beta/Q)
 + 4\beta^2 (\exp(\omega\delta_i/Q)-1)}{8\omega^3\beta^2} &
 \frac{Q \sin^2(\omega\delta_i\beta/Q)}{2\omega^2\beta^2}\\
 \frac{Q\sin^2(\omega\delta_i\beta/Q)}{2\omega^2\beta^2}&
  \frac{(\cos(2\omega\delta_i\beta/Q)-1) + 2\beta \sin(2\omega\delta_i\beta/Q) + 4\beta^2(\exp(\omega\delta_i/Q)-1)}{8\omega\beta^2}\\
\end{array}
\right)
\end{align}

\subsection{DSHO with parameters $\omega$ and $Q$. Case $Q<1/2$.}

Define $\beta=\sqrt{1/4-Q^2}$. The expressions for $\bm{\Phi}$ and $\bm{Q}$ correspond to those of the case $Q\geq1/2$ using the fact that $\cos(ix)=\cosh(x)$ and $\sin(ix)=i\sinh(x)$.

\begin{equation}
\bm{\Phi}_{i-1} = \exp(-\omega \delta_i/2Q)
\left(
\begin{array}{cc}
 \cosh(\omega\beta\delta_i/Q) + \sinh(\omega\beta\delta_i/Q)/2\beta &
 Q\sinh(\omega\beta\delta_i/Q)/\omega\beta\\
  -Q\omega\sinh(\omega\delta_i\beta/Q)/\beta &
 \cosh(\omega\delta_i\beta/Q) - \sinh(\omega\delta_i\beta/Q)/2\beta \\
\end{array}
\right)
\end{equation}

\begin{align}
\bm{Q}_{i-1} =&
Q\exp(-\omega\delta_i/Q)\times\\
& \left(
\begin{array}{cc}
 \frac{(1-\cosh(2\omega\delta_i\beta/Q)) - 2\beta \sinh(2\omega\delta_i\beta/Q)
 + 4\beta^2 (\exp(\omega\delta_i/Q)-1)}{8\omega^3\beta^2} &
 \frac{Q \sinh^2(\omega\delta_i\beta/Q)}{2\omega^2\beta^2}\\
 \frac{Q\sinh^2(\omega\delta_i\beta/Q)}{2\omega^2\beta^2}&
  \frac{(1-\cosh(2\omega\delta_i\beta/Q)) + 2\beta \sinh(2\omega\delta_i\beta/Q) + 4\beta^2(\exp(\omega\delta_i/Q)-1)}{8\omega\beta^2}\\
\end{array}
\right)
\end{align}

Note that the $\bm{\Phi}$ and $\bm{Q}$ matrices for the DSHO process have well-defined limits as $Q\rightarrow 0.5^+$ and $Q\rightarrow 0.5^-$, and are given by the corresponding matrices for the Mat\'ern $\nu=3/2$ process. This follows from the fact that $\lim_{x\rightarrow 0} \sin(x)/x = \lim_{x\rightarrow 0} \sinh(x)/x = 1$.

\bibliography{gp-state.bib}
\end{document}